\documentclass[english,aps,superscriptaddress]{revtex4}
\usepackage[T1]{fontenc}
\usepackage[latin9]{inputenc}
\usepackage{color}
\usepackage{babel}
\usepackage{verbatim}
\usepackage{amsmath}
\usepackage{amssymb}
\usepackage{graphicx}
\usepackage[unicode=true,pdfusetitle,
 bookmarks=true,bookmarksnumbered=false,bookmarksopen=false,
 breaklinks=false,pdfborder={0 0 1},backref=false,colorlinks=false]
 {hyperref}

\makeatletter

\providecommand{\tabularnewline}{\\}

\@ifundefined{textcolor}{}
{%
 \definecolor{BLACK}{gray}{0}
 \definecolor{WHITE}{gray}{1}
 \definecolor{RED}{rgb}{1,0,0}
 \definecolor{GREEN}{rgb}{0,1,0}
 \definecolor{BLUE}{rgb}{0,0,1}
 \definecolor{CYAN}{cmyk}{1,0,0,0}
 \definecolor{MAGENTA}{cmyk}{0,1,0,0}
 \definecolor{YELLOW}{cmyk}{0,0,1,0}
}

\makeatother

\begin{document}
\title{A solvable model for spin polarizations with flow-momentum correspondence}
\author{Anum Arslan}
\email{anumramay@mail.ustc.edu.cn}

\affiliation{Department of Modern Physics and Anhui Center for fundamental sciences
in theoretical physics, University of Science and Tchenology of China,
Hefei, Anhui 230026, China}
\author{Wen-Bo Dong}
\email{wenba@mail.ustc.edu.cn}

\affiliation{Department of Modern Physics and Anhui Center for fundamental sciences
in theoretical physics, University of Science and Tchenology of China,
Hefei, Anhui 230026, China}
\author{Guo-Liang Ma}
\email{glma@fudan.edu.cn}

\affiliation{Key Laboratory of Nuclear Physics and Ion-beam Application (MOE),
Institute of Modern Physics, Fudan University, Shanghai, 200433, China}
\affiliation{Shanghai Research Center for Theoretical Nuclear Physics, NSFC and
Fudan University, Shanghai, 200438, China}
\author{Shi Pu}
\email{shipu@ustc.edu.cn}

\affiliation{Department of Modern Physics and Anhui Center for fundamental sciences
in theoretical physics, University of Science and Tchenology of China,
Hefei, Anhui 230026, China}
\author{Qun Wang}
\email{qunwang@ustc.edu.cn}

\affiliation{Department of Modern Physics and Anhui Center for Fundamental Sciences
in Theoretical Physics, University of Science and Tchenology of China,
Hefei, Anhui 230026, China}
\affiliation{Department of Physics, McGill University, Montreal, Quebec H3A 2T8,
Canada}
\begin{abstract}
We present an analytically solvable model based on the blast-wave
picture of heavy-ion collisions with flow-momentum correspondence.
It can describe the key features of spin polarizations in heavy-ion
collisions. With the analytical solution, we can clearly show that
the spin polarization with respect to the reaction plane is governed
by the directed flow, while the spin polarization along the beam direction
is governed by the ellipticity in flow and in transverse emission
area. There is a symmetry between the contribution from the vorticity
and from the shear stress tensor due to the flow-momentum correspondence.
The solution can be improved systematically by perturbation method.

\end{abstract}
\maketitle

\section{Introduction}

In non-central heavy-ion collisions, a large orbital angular momentum
of two colliding nuclei can be partially converted into the spin polarization
of particles in the final state \citep{Liang:2004ph}. This effect
is called the global spin polarization \citep{Liang:2004ph,Voloshin:2004ha,Gao:2007bc,Betz:2007kg,Becattini:2007sr}
since it is with respect to the reaction plane formed by the impact
parameter and the beam direction which is the same for all particles
in one event. In contrast, in hadron-hadron or electron-positron collisions
the particle's spin polarization is with respect to production planes
formed by the particle's momentum and the beam direction which are
different for particles with different momenta in one event. The global
spin polarization of $\Lambda$ hyperons was measured by STAR collaboration
in Au+Au collisions from 3 to 200 GeV \citep{STAR:2007ccu,STAR:2017ckg,STAR:2018gyt},
by HADES collaboration in Au+Au and Ag+Ag collisions at 2.42-2.55
GeV \citep{HADES:2022enx}, and by ALICE collaboration in Pb+Pb collisions
at 5.02 TeV \citep{ALICE:2019onw}. The global spin polarization can
be described through simulations in hydrodynamic models and transport
models such as AMPT and URQMD models \citep{Xia:2018tes,Karpenko:2016jyx,Sun:2017xhx,Li:2017slc,Wei:2018zfb,Vitiuk:2019rfv,Becattini:2020ngo,Fu:2020oxj,Ryu:2021lnx}
using Cooper-Frye formula \citep{Becattini:2013fla,Fang:2016vpj}.
For recent reviews on global spin polarizations, we refer readers
to Refs. \citep{Wang:2017jpl,Florkowski:2018fap,Huang:2020dtn,Gao:2020lxh,Gao:2020vbh,Liu:2020ymh,Becattini:2022zvf,Hidaka:2022dmn,Becattini:2024uha}. 


It was later proposed in the hydrodynamic model that the spin polarization
along the beam direction should behaves as $P^{z}\sim-\sin(2\phi_{p})$
where $\phi_{p}$ is the azimuthal angle of the hyperon's momentum
in the transverse plane relative to that of the reaction plane \citep{Becattini:2017gcx}.
This prediction was confirmed by the transport model \citep{Xia:2018tes}.
The SATR collaboration measured $P^{z}$ but found a sign difference
from the theorectical prediction \citep{STAR:2019erd}, which is called
the ``sign puzzle''. The first theoretical attempt to explain experimental
data was made by the authors of Refs.~\citep{Wu:2019eyi,Wu:2020yiz}
who found that the temperature vorticity can qualitatively explain
the spin polarization along the beam direction. It was later found
that a contribution from the shear stress tensor~\citep{Fu:2021pok,Becattini:2021iol,Yi:2021ryh,Florkowski:2021xvy,Wagner:2022gza,Wu:2022mkr}
can give the correct sign in the longitudinal polarization. It was
recently proposed that the projected thermal vorticity and the dissipative
correction in a thermal model can describe the behavior of the longitudinal
polarization \citep{Banerjee:2024xnd}. All these theoretical models
show that the global equilibrium has not been reached and off-equilibrium
effects have to be considered \citep{Sheng:2021kfc}.


Actually the blast wave model was used in Ref. \citep{STAR:2019erd}
by the STAR collaboration to explain the data for the spin polarization
along the beam direction \citep{Voloshin:2017kqp,Niida:2018hfw}.
As is well known, the blast-wave model can describe particle's momentum
spectra on the freeze-out hypersurface \citep{Retiere:2003kf}. The
assumption or picture of the blast wave model is that in high energy
hevy-ion collisions the hot and dense matter becomes a thermal source
for particle emission. The released particles are assumed to freeze
out freely without interacting with other particles. The system is
assumed to have longitudinal boost invariance. The transverse rapidity
depends linearly on the freeze-out radius, so the flow velocity is
smaller near the center and is larger near the edge of the emission
region. In this form, the velocity field looks like a blast wave.
Different blast-wave models have been used in describing particle
momentum spectra \citep{STAR:2008med,Ristea:2013ara,Waqas:2019mjp,Chen:2020zuw,Wang:2020wvu,Zhang:2014jug}
and collective flows \citep{PHENIX:2015jaj,Nara:2022kbb} in heavy-ion
collisions at various collision energies. 


The spin polarization along the beam direction was explained by the
blast-wave picture \citep{Siemens:1978pb,Lee:1990sk,Schnedermann:1993ws,Huovinen:2001cy}
under a naive approximation \citep{STAR:2019erd,Voloshin:2017kqp}:
$P^{z}\approx\omega^{z}/2$. Then it is quite natural to understand
the pattern $P^{z}\sim\sin(2\phi)$ seen in the experiment from the
elliptic flow $v_{2}$. As illustrated in Fig. \ref{fig:overlap-region},
the radial flow velocity in the transverse plane has the form $\mathbf{v}\sim\mathbf{e}_{r}v_{r}\left[1+v_{2}\cos(2\phi)\right]$,
where $v_{r}$ is the radial flow velocity and $\mathbf{e}_{r}=(\cos\phi,\sin\phi)$
is the radial direction in the transverse plane. With such a profile
of the flow velocity, we immediately obtain $P^{z}\sim\partial_{x}v^{y}-\partial_{y}v^{x}\sim(1/r)v_{2}v_{r}\sin(2\phi)$
which is proportional to $v_{2}$. However, the blast-wave model used
in Ref. \citep{STAR:2019erd,Voloshin:2017kqp} fails to reproduce
the global spin polarization because the flow velocity is boost invariant
and cannot describe a rotation along $y$ direction. Furthermore,
the approximation $P^{z}\approx\omega^{z}/2$ is non-relativistic
and does not incldue the relativistic effect which might change the
pattern seen in the experiment.

Inspired by the potential of the blast wave picture in describing
the spin polarization along the beam direction, we will perform a
comprehensive analysis of global and local spin polarization using
the modified (or improved) blast wave model (MBWM) \citep{Retiere:2003kf,Jaiswal:2015saa,Yang:2016rnw,Yang:2020oig,Yang:2022yxa}.
The essential part of the MBWM is the profile of the flow velocity
$u^{\mu}$, a four-vector as a function of the space time rapidity
$\eta$ in the longitudinal direction, the transverse rapidity $\rho$,
and the azimuthal angle $\phi_{b}$ in the transverse plane. To describe
the global spin polarization with respect to the reaction plane, we
introduce into the transverse rapidity $\rho$ a term $\alpha_{1}\eta\cos\phi_{b}$
related to the directed flow $v_{1}$, which slightly breaks the boost
invariance and leads to a rotation of the system along $-y$ direction.
There is also a term $\rho_{2}\cos(2\phi_{b})$ in $\rho$ related
to the elliptic flow $v_{2}$. With this profile of $u^{\mu}$ and
under flow-momentum correspondence, we can derive the analytical formula
for $P^{z}$ and $P^{y}$ as functions of $\phi_{p}$ and $p_{T}$
which capture the key patterns observed in experiments. To our knowledge,
this is the first analytically solvable model for spin polarizations
in heavy-ion collisions. 


In this paper, we adopt the following notations: $g^{\mu\nu}=\text{diag}\left(1,-1,-1,-1\right)$
where $\mu,\nu=0,1,2,3$. Levi-Civita symbol is defined as $\epsilon^{0123}=-\epsilon_{0123}=1$,
$\hbar=k_{B}=1$, $x^{\mu}=\left(x^{0},\mathbf{x}\right)$ and $u\cdot p=u^{\mu}p_{\mu}$.
The summation of repeated indices is implied if not stated explicitly.


\section{Description of the model \label{sec:modefied-blast-wave-model}}

\begin{figure}
\includegraphics[scale=0.6]{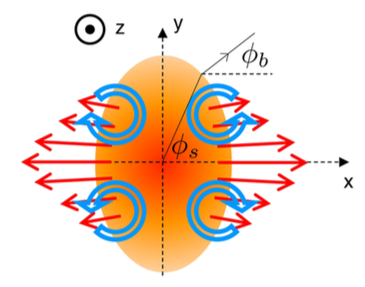}

\caption{The overlap region of two colliding nuclei. The beam are along $\pm z$
driection. Here the reaction plane is formed by the $x$ and $z$
directions. The figure is taken from Ref. \citep{STAR:2019erd} \label{fig:overlap-region}}

\end{figure}

Our solvable model is based on the modified blast-wave model to which
we will give a brief introduction in this section. The overlap region
of two colliding nuclei in the transverse plane ($xy$ plane) is shown
in Fig. \ref{fig:overlap-region} with the nucleus at $x=\pm b/2$
going in the $\pm z$ direction. The perpendicular direction of the
reaction plane is then along the $y$ direction.


The QGP can be approximated as a longitudinally boost-invariant system,
thus it is natural to use the proper time $\tau$ and the space-time
rapidity $\eta$ as variables to replace $t$ and $z$. Accordingly,
the particle's momentum in longitudinal direction can also be described
by the transverse mass $m_{T}$ and the momentum rapidity $Y$. These
variables are defined as 
\begin{align}
\tau= & \sqrt{t^{2}-z^{2}},\;\eta=\frac{1}{2}\ln\frac{t+z}{t-z},\nonumber \\
m_{T}= & \sqrt{m^{2}+p_{T}^{2}},\;Y=\frac{1}{2}\ln\frac{E_{p}+p_{z}}{E_{p}-p_{z}},
\end{align}
where $E_{p}=\sqrt{|\mathbf{p}|^{2}+m^{2}}$, $\mathbf{p}\equiv(p_{x},p_{y},p_{z})$,
and $\mathbf{p}_{T}=(p_{x},p_{y})$ with $p_{T}\equiv|\mathbf{p}_{T}|$.
Therefore the flow four-velocity and the particle's four-momentum
can be parameterized as 
\begin{align}
u^{\mu}(x)= & \left(\cosh\eta\cosh\rho,\sinh\rho\cos\phi_{b},\sinh\rho\sin\phi_{b},\sinh\eta\cosh\rho\right),\label{eq:flow-velocity}\\
p^{\mu}= & \left(m_{T}\cosh Y,p_{T}\cos\phi_{p},p_{T}\sin\phi_{p},m_{T}\sinh Y\right).\label{eq:particle-momentum}
\end{align}
Here the transverse expansion of the fireball \citep{Lee:1990sk,Huovinen:2001cy,Retiere:2003kf}
is decribed by the transverse rapidity $\rho$ as a function of $\widetilde{r}$
(normalized transverse radius), $\phi_{b}$ (azimuthal angle of the
flow velocity in transverse plane) and $\eta$ as follows 
\begin{eqnarray}
\rho\left(r,\phi_{s},\eta\right) & = & \widetilde{r}\left[\rho_{0}+\rho_{1}(\eta)\cos(\phi_{b})+\rho_{2}\cos(2\phi_{b})\right],\label{eq:rho-1}
\end{eqnarray}
where $\phi_{b}$ is a function of the azimuthal angle $\phi_{s}$
in transverse plane in coordinate space (the function will be defined
later), $\rho_{0}$ characterizes the mean transverse rapidity of
the source element, $\rho_{1}(\eta)=\alpha_{1}\eta$ and $\rho_{2}$
describe the azimuthal anisotropy of the transverse rapidity. In Eq.
(\ref{eq:rho-1}) $\widetilde{r}$ is defined as 
\begin{align}
\widetilde{r}= & \sqrt{\frac{(r\cos\phi_{s})^{2}}{R_{x}^{2}}+\frac{(r\sin\phi_{s})^{2}}{R_{y}^{2}}}\nonumber \\
\approx & \frac{r}{R}\left[1+\frac{1}{2}\epsilon\cos(2\phi_{s})\right],
\end{align}
where $R_{x}$ and $R_{y}$ are effective radii of the elliptic source
in $x$- and $y$-direction respectively, and in the second line we
have used the approximation $\epsilon\equiv R_{y}-R_{x}\ll R\equiv(R_{x}+R_{y})/2$
which works very well in describing data \citep{Retiere:2003kf}.
The functional relation between $\phi_{b}$ and $\phi_{s}$ is 
\begin{equation}
\tan\phi_{b}=\frac{R_{x}^{2}}{R_{y}^{2}}\tan\phi_{s}\approx(1-2\epsilon)\tan\phi_{s},
\end{equation}
as shown in Fig. \ref{fig:overlap-region}. In this paper we will
use the ordering of parameters
\begin{equation}
\alpha_{1}\sim\rho_{2}\sim\epsilon\ll\rho_{0}.
\end{equation}
We denote $\alpha_{1}$, $\rho_{2}$ and $\epsilon$ are $O(\epsilon)$
quantities, while $\rho_{0}$ is an $O(1)$ quantity, so $\alpha_{1}$,
$\rho_{2}$ and $\epsilon$ can be treated as perturbations relative
to $\rho_{0}$.


The particle's distribution function in phase space $f(x,p)$ is assumed
to follow the Boltzmann distribution under the condition $\beta p\cdot u\gg1$
($\beta=1/T$ is the inverse temperature), 
\begin{align}
f(x,p)\equiv & f(p\cdot u)=\exp(-\beta p\cdot u)\nonumber \\
= & \exp\left\{ -\beta\left[m_{T}\cosh\rho\cosh(\eta-Y)-p_{T}\sinh\rho\cos(\phi_{b}-\phi_{p})\right]\right\} ,\label{eq:distribution}
\end{align}
following Eqs. (\ref{eq:flow-velocity}) and (\ref{eq:particle-momentum}).
One can see that the distribution (\ref{eq:distribution}) reaches
a maximum when $\eta\approx Y$ and $\phi_{b}\approx\phi_{p}$, i.e.
the spacetime and momentum rapidities are equal and the flow and momentum
azimuthal angles are equal. If we set $\eta=Y$ and $\phi_{b}=\phi_{p}$,
the distribution reaches a maximum at $p_{T}/m_{T}=\tanh\rho$, i.e.
the transverse momentum rapidity is equal to the transverse flow rapidity.
These conditions are called the flow-momentum correspondence in fireball's
expansion.


Physical observables can be computed on the freeze-out hypersurface
by 
\begin{equation}
\left\langle O(p)\right\rangle =\frac{\int d^{4}xO(x,p)S(x,p)}{\int d^{4}xS(x,p)}.\label{eq:weight-ob}
\end{equation}
Here $O(x,p)$ is the phyiscal quantity in phase space corresponding
to the observable, and the emission function $S(x,p)$ represents
the probability of emitting a particle with the momentum $p$ at the
space-time $x$ and thus defines a freeze-out hyper-surface for particle
emission at the freeze-out temperature $T=T_{f}$, 
\begin{equation}
S(x,p)=m_{T}\cosh(\eta-Y)\delta(\tau-\tau_{f})\Theta(R-r)f(x,p),\label{eq:emission-function}
\end{equation}
where $\tau_{f}$ is the freeze-out proper time and $\Theta(x)$ is
the Heaviside step function to require that all particles be emitted
within the sphere volume of the radius $R$. The momentum integrated
observables can be obtained by integration over all components of
the on-shell momentum
\begin{equation}
\left\langle O\right\rangle =\frac{\int d^{4}xd^{3}\mathbf{p}E_{p}^{-1}O(x,p)S(x,p)}{\int d^{4}xd^{3}\mathbf{p}E_{p}^{-1}S(x,p)}.\label{eq:int-weight-ob}
\end{equation}
The integral elements of space-time and on-shell momentum are 
\begin{align}
d^{4}x= & \tau rd\tau d\eta drd\phi_{s},\nonumber \\
\frac{d^{3}\mathbf{p}}{E_{p}}= & p_{T}dp_{T}dYd\phi_{p}.
\end{align}
The partially integrated observables can also be obtained by integration
over some components of the on-shell momentum. For example, if we
look at the $p_{T}$ and $\phi_{p}$ dependence of the observable,
we can leave $p_{T}$ and $\phi_{p}$ out of the momentum integral
\begin{align}
\left\langle O\right\rangle (p_{T})= & \frac{\int d^{4}xdYd\phi_{p}O(x,p)S(x,p)}{\int d^{4}xdYd\phi_{p}S(x,p)},\nonumber \\
\left\langle O\right\rangle (\phi_{p})= & \frac{\int d^{4}xdp_{T}dYp_{T}O(x,p)S(x,p)}{\int d^{4}xdp_{T}dYp_{T}S(x,p)}.\label{eq:pt-phi-p}
\end{align}
We will use the above formula to compute the polarization in the global
OAM and beam directions as functions of $p_{T}$ or $\phi_{p}$.


\section{Collective flows and spin vectors: analytical results}

\label{sec:Collective-flows}With Eqs. (\ref{eq:flow-velocity}) and
(\ref{eq:particle-momentum}) for $u^{\mu}$ and $p^{\mu}$ as well
as Eqs. (\ref{eq:distribution}) and (\ref{eq:emission-function})
for $f(p\cdot u)$ and $S(x,p)$, we can calculate the directed flow
$v_{1}$ as a function of $Y$ and the elliptic flow as a function
of $p_{T}$ as 
\begin{align}
v_{1}(Y)= & \frac{\int d^{4}x\int dp_{T}d\phi_{p}p_{T}\;\cos(\phi_{p})S(x,p)}{\int d^{4}x\int dp_{T}d\phi_{p}p_{T}\;S(x,p)},\nonumber \\
v_{2}(p_{T})= & \frac{\int d^{4}x\int dYd\phi_{p}\;\cos(2\phi_{p})S(x,p)}{\int d^{4}x\int dYd\phi_{p}\;S(x,p)}.
\end{align}
In the leading order of the flow-momentum correspondence with $\eta=Y$
and $\phi_{b}=\phi_{p}$, we can obtain analytical results for $v_{1}(Y)$
and $v_{2}(p_{T})$ at the central rapidity $Y\approx0$, 
\begin{align}
v_{1}(Y)= & \alpha_{1}Y\frac{\beta}{2R}\frac{N_{v1}}{N_{v0}},\nonumber \\
v_{2}(p_{T})= & \left(\rho_{2}+\frac{1}{2}\epsilon\rho_{0}\right)\frac{\beta}{2R}\frac{N_{v2}(p_{T})}{N_{v0}(p_{T})}+\epsilon\frac{N_{v2}^{\epsilon}(p_{T})}{N_{v0}(p_{T})},\label{eq:v1-v2}
\end{align}
where we have implied $\beta\equiv1/T_{f}$, $N_{v0}$ and $N_{v1}$
are constants in $v_{1}(Y)$ which are defined through $p_{T}$ functions
$N_{v0}(p_{T})$ and $N_{v1}(p_{T})$, and $N_{v2}(p_{T})$ and $N_{v2}^{\epsilon}(p_{T})$
as well as $N_{v0}(p_{T})$ are $p_{T}$ functions used in $v_{2}(p_{T})$.
All these quantities are defined as 
\begin{align}
N_{0}= & \int_{0}^{p_{T}^{\mathrm{max}}}dp_{T}p_{T}\;N_{0}(p_{T}),\nonumber \\
N_{v1}= & \int_{0}^{p_{T}^{\mathrm{max}}}dp_{T}p_{T}N_{v1}(p_{T}),\nonumber \\
N_{0}(p_{T})= & \int_{0}^{R}dr\;rm_{T}\;K_{1}(\beta m_{T}\cosh\bar{\rho})I_{0}(\beta p_{T}\sinh\bar{\rho}),\nonumber \\
N_{v1}(p_{T})= & \int_{0}^{R}dr\;r^{2}m_{T}\left[m_{T}\sinh\bar{\rho}K_{1}^{\prime}(\beta m_{T}\cosh\bar{\rho})I_{1}(\beta p_{T}\sinh\bar{\rho})\right.\nonumber \\
 & \left.+p_{T}\cosh\bar{\rho}K_{1}(\beta m_{T}\cosh\bar{\rho})I_{1}^{\prime}(\beta p_{T}\sinh\bar{\rho})\right],\nonumber \\
N_{v2}(p_{T})= & \int_{0}^{R}dr\;r^{2}m_{T}\left[m_{T}\sinh\bar{\rho}K_{1}^{\prime}(\beta m_{T}\cosh\bar{\rho})I_{2}(\beta p_{T}\sinh\bar{\rho})\right.\nonumber \\
 & \left.+p_{T}\cosh\bar{\rho}K_{1}(\beta m_{T}\cosh\bar{\rho})I_{2}^{\prime}(\beta p_{T}\sinh\bar{\rho})\right],\nonumber \\
N_{v2}^{\epsilon}(p_{T})= & \int_{0}^{R}dr\;rm_{T}\;K_{1}(\beta m_{T}\cosh\bar{\rho})I_{2}(\beta p_{T}\sinh\bar{\rho}),\label{eq:nv012}
\end{align}
where $I_{0,1,2}(x)$ and $K_{1}(x)$ are modified Bessel functions
of the first and second kind respectively, $I_{1,2}^{\prime}(x)=dI_{1,2}(x)/dx$
and $K_{1}^{\prime}(x)=dK_{1}(x)/dx$ are their derivatives, and $\bar{\rho}\equiv(r/R)\rho_{0}$
is just the $\rho_{0}$ part of $\rho$ in (\ref{eq:rho-1}).


In this paper we aim to calculate the spin polarization of spin-1/2
particles. The spin vectors are defined as 
\begin{align}
\hat{P}_{\omega}^{\mu}= & -\frac{1}{4m}\epsilon^{\mu\nu\sigma\tau}(1-f)\omega_{\nu\sigma}p_{\tau},\label{eq:VIP}\\
\hat{P}_{\xi}^{\mu}= & -\frac{1}{2m}\epsilon^{\mu\nu\sigma\tau}(1-f)\frac{p_{\tau}p^{\rho}}{E_{p}}\hat{t}_{\nu}\xi_{\rho\sigma},\label{eq:SIP}
\end{align}
where the vector $\hat{t}$ is given by $\hat{t}^{\mu}=(1,0,0,0)$
in the laboratory frame, $\omega^{\mu\nu}$ and $\xi^{\mu\nu}$ denote
the thermal vorticity and thermal shear stress tensor respectively
which are defined as 
\begin{align}
\omega^{\mu\nu}= & -\frac{1}{2}\left[\partial^{\mu}\left(\beta u^{\nu}\right)-\partial^{\nu}\left(\beta u^{\mu}\right)\right],\nonumber \\
\xi^{\mu\nu}= & \frac{1}{2}\left[\partial^{\mu}\left(\beta u^{\nu}\right)+\partial^{\nu}\left(\beta u^{\mu}\right)\right].
\end{align}
Note that the polarization vector defined in Eq. (\ref{eq:VIP}) is
different from Ref. \citep{STAR:2019erd}, in which $\hat{P}_{\omega}^{z}$
was assumed to be simply proportional to $\omega^{xy}$.


With Eqs. (\ref{eq:flow-velocity}) and (\ref{eq:particle-momentum})
for $u^{\mu}$ and $p^{\mu}$, we can calculate spin vectors $\hat{P}_{\omega}^{i}$
and $\hat{P}_{\xi}^{i}$ (for $i=y,z$) as functions of $\phi_{p}$
following Eqs. (\ref{eq:VIP}) and (\ref{eq:SIP}). In the leading
order of the flow-momentum correspondence with $\eta=Y=0$ and $\phi_{b}=\phi_{p}$,
the analytical results for $\hat{P}_{\omega}^{i}$ and $\hat{P}_{\xi}^{i}$
can be obtained to $O(\epsilon)$, 
\begin{align}
\hat{P}_{\omega}^{y}\approx & \frac{1}{4mT\tau}\left[\alpha_{1}\frac{r}{R}\left(m_{T}\cosh\rho-p_{T}\sinh\rho\right)\cos^{2}\phi_{p}\right.\nonumber \\
 & \left.+C_{\omega}^{\eta}(p_{T},\phi_{p})\partial_{\eta=0}\ln T\right]\nonumber \\
\hat{P}_{\xi}^{y}\approx & \frac{1}{4mT\tau}\frac{p_{T}}{m_{T}}\left[\alpha_{1}\frac{r}{R}\left(p_{T}\cosh\rho-m_{T}\sinh\rho\right)\cos^{2}\phi_{p}\right.\nonumber \\
 & \left.+C_{\xi}^{\eta}(p_{T},\phi_{p})\partial_{\eta=0}\ln T\right],\nonumber \\
\hat{P}_{\omega}^{z}\approx & \frac{1}{4mT}\left[2\rho_{2}\frac{1}{R}\left(m_{T}\cosh\rho-p_{T}\sinh\rho\right)\sin(2\phi_{p})-\epsilon\frac{1}{r}m_{T}\sinh\rho\sin(2\phi_{p})\right.\nonumber \\
 & \left.+C_{\omega}^{\phi}(p_{T},\phi_{p})\partial_{\phi_{s}}\ln T+C_{\omega}^{r}(p_{T},\phi_{p})\partial_{r}\ln T\right],\nonumber \\
\hat{P}_{\xi}^{z}\approx & \frac{1}{4mT}\frac{p_{T}}{m_{T}}\left[2\rho_{2}\frac{1}{R}\left(p_{T}\cosh\rho-m_{T}\sinh\rho\right)\sin(2\phi_{p})+\epsilon\frac{1}{r}p_{T}\sinh\rho\sin(2\phi_{p})\right.\nonumber \\
 & \left.+C_{\xi}^{\phi}(p_{T},\phi_{p})\partial_{\phi_{s}}\ln T+C_{\xi}^{r}(p_{T},\phi_{p})\partial_{r}\ln T\right],\label{eq:un-av-py-pz}
\end{align}
where $C_{\omega}^{i}$ and $C_{\xi}^{i}$ ($i=\eta,\phi,r$) are
coefficients of $\partial_{i}\ln T$ (temperature gradients) as functions
of $p_{T}$ and $\phi_{p}$, and we have made the approximation $1-f(p\cdot u)\approx1$
for under the condition $\beta p\cdot u\gg1$. We see in Eq. (\ref{eq:un-av-py-pz})
that there are two parts in $\hat{P}_{\omega}^{i}$ and $\hat{P}_{\xi}^{i}$
($i=y,z$): the kinetic part and temperature gradient part. The latter
involves only gradients of $T$ in $\eta$, $\phi$ and $r$ directions
and will be vanishing once we take averages of these quantities on
the freeze-out hypersurface defined by $\tau=\tau_{f}$ at $T=T_{f}$
(the direction of $\tau$ is perpendicular to those of $\phi_{s}$
and $\eta$) as in Eq. (\ref{eq:emission-function}). So we can safely
drop the temperature gradient part.


The analytical expressions in (\ref{eq:un-av-py-pz}) have some good
features and symmetries. (a) $\hat{P}_{\omega}^{y}$ and $\hat{P}_{\xi}^{y}$
are proportional to $\cos^{2}\phi_{p}$ and driven by the directed
flow $v_{1}$, while $\hat{P}_{\omega}^{z}$ and $\hat{P}_{\xi}^{z}$
are proportional to $\sin(2\phi_{p})$ and driven by the elliptic
flow $v_{2}$ and ellipticity parameter $\epsilon$ of the transverse
emission region. (b) The coefficients as functions of $\rho$ are
different between $\hat{P}_{\omega}^{i}$ and $\hat{P}_{\xi}^{i}$
but closely connected with each other. The collective flow parts of
$\hat{P}_{\omega}^{y}$ and $\hat{P}_{\omega}^{z}$ ($\alpha_{1}$
and $\rho_{2}$ terms) are proportional to $m_{T}\cosh\rho-p_{T}\sinh\rho$
which is positive definite, while those of $\hat{P}_{\xi}^{y}$ and
$\hat{P}_{\xi}^{z}$ ($\alpha_{1}$ and $\rho_{2}$ terms respectively)
are proportional to $p_{T}\cosh\rho-m_{T}\sinh\rho$ which is vanishing
when $p_{T}/m_{T}=\tanh\rho$, i.e. the transverse momentum rapidity
is equal to the transverse flow rapidity, a condition of the flow-momentum
correspondence in transverse expansion. So for $\alpha_{1}$ and $\rho_{2}$
terms, both $P^{y}$ and $P^{z}$ are dominated by the kinetic vorticity
but not from the shear stress tensor, when the flow-momentum correspondence
is implemented. However, in our calculation of the spin polarization,
we implement a partial flow-momentum correspondence with $\eta=Y$
and $\phi_{b}=\phi_{p}$ but not $p_{T}/m_{T}=\tanh\rho$, so the
contribution from the shear stress tensor ($\alpha_{1}$ and $\rho_{2}$
terms) is non-vanishing.


We can insert the quantities in (\ref{eq:un-av-py-pz}) into the second
line of (\ref{eq:pt-phi-p}) to compute the average values $P^{y}$
and $P^{z}$ as functions of $\phi_{p}$ on the freeze-out hyer-surface
defined in Eq. (\ref{eq:emission-function}),
\begin{align}
P^{y}(\phi_{p})= & \left\langle \hat{P}_{\omega}^{y}+\hat{P}_{\xi}^{y}\right\rangle (\phi_{p})\nonumber \\
\approx & \alpha_{1}\frac{1}{4mT_{f}\tau_{f}R}\frac{1}{N_{0}}\left[N_{1}(2,1,2)+N_{1}(2,3,0)-2N_{2}(2,2,1)\right]\cos^{2}\phi_{p},\nonumber \\
P^{z}(\phi_{p})= & \left\langle \hat{P}_{\omega}^{z}+\hat{P}_{\xi}^{z}\right\rangle (\phi_{p})\nonumber \\
\approx & \rho_{2}\frac{1}{2mT_{f}R}\frac{1}{N_{0}}\left[N_{1}(1,1,2)+N_{1}(1,3,0)-2N_{2}(1,2,1)\right]\sin(2\phi_{p})\nonumber \\
 & -\epsilon\frac{1}{4mT_{f}}\frac{1}{N_{0}}\left[N_{2}(0,1,2)-N_{2}(0,3,0)\right]\sin(2\phi_{p}),\label{eq:py-pz-phi}
\end{align}
where the normalization constant $N_{0}$ given by the first line
of Eq. (\ref{eq:nv012}), and $N_{1,2}(n_{1},n_{2},n_{3})$ involve
following integrals over $p_{T}$ and $r$ 
\begin{align}
N_{1}(n_{1},n_{2},n_{3})= & \int_{0}^{p_{T}^{\mathrm{max}}}dp_{T}\int dr\;r^{n_{1}}p_{T}^{n_{2}}m_{T}^{n_{3}}\;\cosh\bar{\rho}\;K_{1}(\beta m_{T}\cosh\bar{\rho})I_{0}(\beta p_{T}\sinh\bar{\rho}),\nonumber \\
N_{2}(n_{1},n_{2},n_{3})= & \int_{0}^{p_{T}^{\mathrm{max}}}dp_{T}\int dr\;r^{n_{1}}p_{T}^{n_{2}}m_{T}^{n_{3}}\;\sinh\bar{\rho}\;K_{1}(\beta m_{T}\cosh\bar{\rho})I_{0}(\beta p_{T}\sinh\bar{\rho}).\label{eq:anal-int}
\end{align}
In Eq. (\ref{eq:anal-int}) we have implied $\beta\equiv1/T_{f}$.
The structure of the collective flow parts ($\alpha_{1}$ and $\rho_{2}$
terms) of $P^{y}$ and $P^{z}$ as functions of $\phi_{p}$ is quite
similar: (a) $P^{y}$ is governed by the directed flow $v_{1}$ and
depends on $\cos^{2}\phi_{p}$, while $P^{z}$ is governed by the
elliptic flow $v_{2}$ and depends on $\sin(2\phi_{p})$; (b) The
prefactors of $\cos^{2}\phi_{p}$ and $\sin(2\phi_{p})$ have the
same structure except the integrals of $P^{y}$ have an additional
factor $r/\tau_{f}$ in the integrands relative to $P^{z}$.


We can also obtain $P^{y}(p_{T})$ and $P_{\sin(2\phi)}^{z}(p_{T})$
by integration over $\phi_{p}$ instead of $p_{T}$, 
\begin{align}
P^{y}(p_{T})= & \left\langle \hat{P}_{\omega}^{y}+\hat{P}_{\xi}^{y}\right\rangle (p_{T})\nonumber \\
\approx & \alpha_{1}\frac{1}{8mT_{f}R\tau_{f}}\frac{1}{N_{0}(p_{T})}\left[N_{p1}(2,0,2)+N_{p1}(2,2,0)-2N_{p2}(2,1,1)\right],\nonumber \\
P_{\sin(2\phi)}^{z}(p_{T})\equiv & \left\langle \left(\hat{P}_{\omega}^{z}+\hat{P}_{\xi}^{z}\right)\sin(2\phi_{p})\right\rangle (p_{T})\nonumber \\
\approx & \rho_{2}\frac{1}{4mT_{f}R}\frac{1}{N_{0}(p_{T})}\left[N_{p1}(1,0,2)+N_{p1}(1,2,0)-2N_{p2}(1,1,1)\right]\nonumber \\
 & -\epsilon\frac{1}{8mT_{f}}\frac{1}{N_{0}(p_{T})}\left[N_{p2}(0,0,2)-N_{p2}(0,2,0)\right],\label{eq:py-pz-pt}
\end{align}
where $N_{0}(p_{T})$ is given in the fourth line of Eq. (\ref{eq:nv012}),
and $N_{p1,p2}(n_{1},n_{2},n_{3})$ are integrals over $r$ depending
on $p_{T}$, 
\begin{align}
N_{p1}(n_{1},n_{2},n_{3})= & \int dr\;r^{n_{1}}p_{T}^{n_{2}}m_{T}^{n_{3}}\;\cosh\bar{\rho}K_{1}(\beta m_{T}\cosh\bar{\rho})I_{0}(\beta p_{T}\sinh\bar{\rho}),\nonumber \\
N_{p2}(n_{1},n_{2},n_{3})= & \int dr\;r^{n_{1}}p_{T}^{n_{2}}m_{T}^{n_{3}}\;\sinh\bar{\rho}K_{1}(\beta m_{T}\cosh\bar{\rho})I_{0}(\beta p_{T}\sinh\bar{\rho}).\label{eq:anal-int-1}
\end{align}
In comparison with Eq. (\ref{eq:py-pz-phi}), the powers of $p_{T}$
in the integrands of $N_{p1}$ and $N_{p2}$ in Eq. (\ref{eq:py-pz-pt})
for $P^{y}(p_{T})$ and $P_{\sin(2\phi)}^{z}(p_{T})$ is less than
those of $N_{1}$ and $N_{2}$ for $P^{y}(\phi_{p})$ and $P^{z}(\phi_{p})$
by 1, respectively.


To obtain the centrality dependence of $P^{y}$ and $P_{\sin(2\phi)}^{z}$,
we need to integrate $N_{0}(p_{T})$ and $N_{p1,p2}(n_{1},n_{2},n_{3})$
over $p_{T}$ in Eq. (\ref{eq:py-pz-pt}) as 
\begin{align}
P^{y}\approx & \alpha_{1}\frac{1}{8mT_{f}R\tau_{f}}\frac{1}{N_{0}}\left[N_{1}(2,1,2)+N_{1}(2,3,0)-2N_{2}(2,2,1)\right],\nonumber \\
P_{\sin(2\phi)}^{z}\approx & \rho_{2}\frac{1}{4mT_{f}R}\frac{1}{N_{0}}\left[N_{1}(1,1,2)+N_{1}(1,3,0)-2N_{2}(1,2,1)\right]\nonumber \\
 & -\epsilon\frac{1}{8mT_{f}}\frac{1}{N_{0}}\left[N_{2}(0,1,2)-N_{2}(0,3,0)\right].\label{eq:py-pz-total}
\end{align}
One can also obtain the above integrated quantities by integration
over $\phi_{p}$ from Eq. (\ref{eq:py-pz-phi}). Note that $P^{y}$
and $P_{\sin(2\phi)}^{z}$ depend on the centrality through relevant
parameters.


\section{Comparison with dada}

In this section, we will calculate collective flows and spin polarizations
using the analytical formula in Sec. \ref{sec:Collective-flows} and
compare with experimental data. 

\subsection{Au+Au collisions at 200 GeV}

\begin{figure}
\centering{}\includegraphics{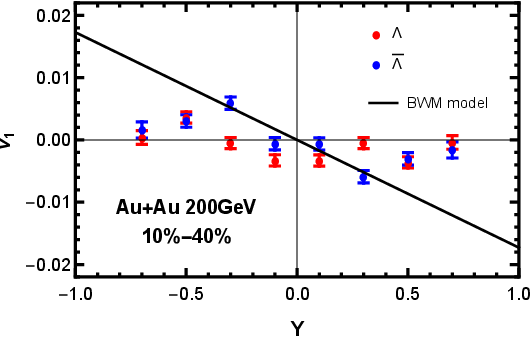}\caption{The result for the directed flow $v_{1}$ of $\Lambda$ and $\overline{\Lambda}$
in Au+Au collisions at 200 GeV. \label{fig:v1}}
\end{figure}

\begin{figure}
\begin{centering}
\includegraphics[scale=0.8]{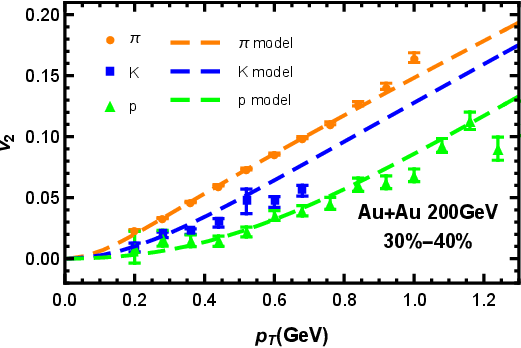} \includegraphics[scale=0.8]{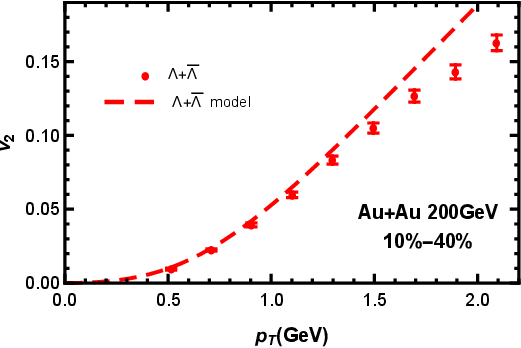}
\par\end{centering}
\caption{The results for the elliptic flows of light particles ($\pi$, $K$,
$p$) (left panel) and $\Lambda+\overline{\Lambda}$ (right panel)
in Au+Au collisions at 200 GeV. The parameters for 20-30\% central
collisions are chosen to fit $\Lambda+\overline{\Lambda}$ data. \label{fig:v2}}
\end{figure}

The parameters that we choose in the solvable model for Au+Au collisions
at $\sqrt{s_{NN}}=200$ GeV and different centralities are listed
in Table \ref{tab:centrality-para}. The freeze-out temperature $T_{f}$
and transverse rapidity parameter $\rho_{0}$ are extracted by fitting
the transverse momentum spectra \citep{STAR:2004jwm,STAR:2008med}.
We fit the data for directed flows of $\Lambda/\overline{\Lambda}$
in 10-40\% central Au+Au collisions by $\alpha_{1}=-0.05$ \citep{STAR:2017okv},
see Fig. \ref{fig:v1}. The parameter $\alpha_{1}$ is assumed to
be independent of the centrality \citep{STAR:2011hyh}. There is a
sizable deviation of the calculated curve for $v_{1}(Y)$ at larger
rapidity from experimental data because of the naive choice of $\rho_{1}=\alpha_{1}\eta$
and approximations made in Sec. \ref{sec:Collective-flows}. Such
a deviation can be improved by introducing a cubic term of $\eta$
into $\rho_{1}$. For elliptic flows, we fit the data for light paticles
and $\Lambda+\overline{\Lambda}$ in different centralities \citep{STAR:2004jwm,STAR:2008ftz}.
As examples, the results for light particles in 30-40\% central collisions
\citep{STAR:2004jwm} and $\Lambda+\overline{\Lambda}$ in 10-40\%
central collisions \citep{STAR:2008ftz} are shown in Fig. \ref{fig:v2},
where we see that the fitted curve agrees well with experimental data.
From Eq. (\ref{eq:v1-v2}), we can find the parameter dependence of
$v_{2}$ easily and extract the $\rho_{2}$ and $\epsilon$ from the
elliptic flow.


\begin{table}
\begin{centering}
\begin{tabular}{|c|c|c|c|c|c|c|c|}
\hline 
centrality & $R$ (fm) & $T$ (MeV) & $\rho_{0}$ & $\rho_{2}$ & $\epsilon$ & $\alpha_{1}$ & $\tau_{f}$(fm/c)\tabularnewline
\hline 
\hline 
10-20\% & 11.5 & 99.5 & 0.982 & 0.023 & 0.05 & $-0.05$ & 7.8\tabularnewline
\hline 
20-30\% & 10.3 & 102 & 0.937 & 0.032 & 0.07 & $-0.05$ & 6.9\tabularnewline
\hline 
30-40\% & 9 & 104 & 0.894 & 0.036 & 0.085 & $-0.05$ & 5.1\tabularnewline
\hline 
40-50\% & 7.8 & 107 & 0.841 & 0.052 & 0.09 & $-0.05$ & 3.3\tabularnewline
\hline 
50-60\% & 7 & 110 & 0.788 & 0.058 & 0.095 & $-0.05$ & 2.6\tabularnewline
\hline 
60-70\% & 6.3 & 116 & 0.707 & 0.068 & 0.11 & $-0.05$ & 2.3\tabularnewline
\hline 
70-80\% & 5.5 & 125 & 0.608 & 0.074 & 0.125 & $-0.05$ & 2.0\tabularnewline
\hline 
\end{tabular}
\par\end{centering}
\caption{Parameters in the solvable model that are used in this paper for Au+Au
collisions at $\sqrt{s_{NN}}=200$ GeV. \label{tab:centrality-para}}
\end{table}

\begin{figure}
\begin{centering}
\includegraphics[scale=0.8]{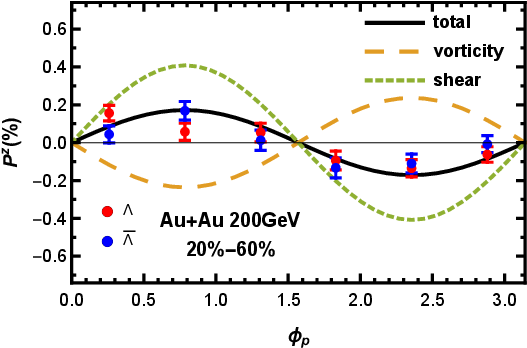} \includegraphics[scale=0.8]{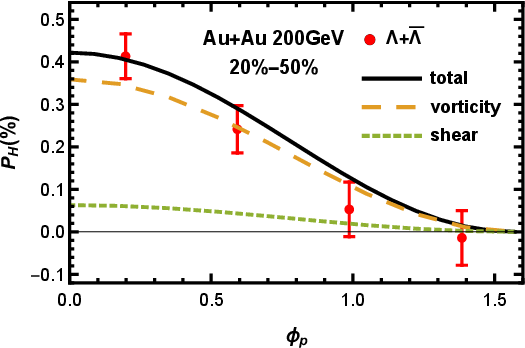}
\par\end{centering}
\caption{The results for $P^{z}$ (right panel) and $P_{H}\equiv-P^{y}$ (left
panel) as functions of $\phi_{p}$ following Eq.(\ref{eq:py-pz-phi})
in Au+Au collisions at 200 GeV. The orange dashed lines and green
dotted lines represent the contributions from the vorticity and shear
stress tensor, respectively. The black solid lines represent the sums
of both contributions. \label{fig:P_phip}}
\end{figure}

The spin polarizations $P^{z}$ and $P_{H}\equiv-P^{y}$ can be calculated
by Eqs. (\ref{eq:py-pz-phi},\ref{eq:py-pz-pt}). In the calculation
we distinguish the contributions from the vorticity and shear stress
tensor as in Eq. (\ref{eq:un-av-py-pz}). The experimental data for
$P^{z}$ are available in 20-60\% centrality Au+Au collisions at $\sqrt{s_{NN}}=200\text{ GeV}$
\citep{STAR:2019erd}, while the $P^{y}$ data are available in 20-50\%
centrality \citep{STAR:2018gyt}. In this paper, the decay parameters
of $\Lambda$ and $\overline{\Lambda}$ are chosen to be $\alpha_{\Lambda}=-\alpha_{\overline{\Lambda}}=0.732$.
The comparision between the model results and experimental data are
shown in Fig. \ref{fig:P_phip}. For $P^{z}$, the shear contribution
is the dominant while the vorticity contribution has an opposite sign
relative to the data, consistent with the results by hydrodynamic
models \citep{Becattini:2021iol,Fu:2021pok,Yi:2021ryh}. This result
is different from Ref. \citep{Voloshin:2017kqp} because a non-relativistic
approximation $P^{z}\approx\omega^{z}/2$ is used there. For $P_{H}$,
we find the vorticity is the dominent source and the shear contribution
is negligible, consistent with the previous discussion in Sec. \ref{sec:Collective-flows}. 

\begin{figure}
\begin{centering}
\includegraphics[scale=0.8]{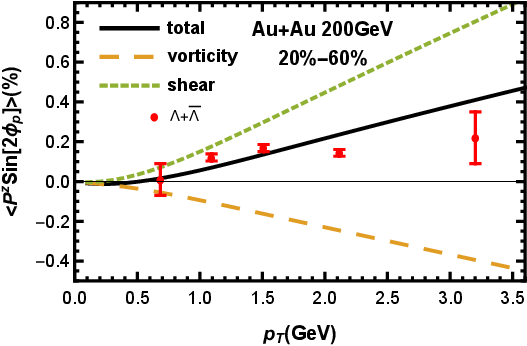} \includegraphics[scale=0.8]{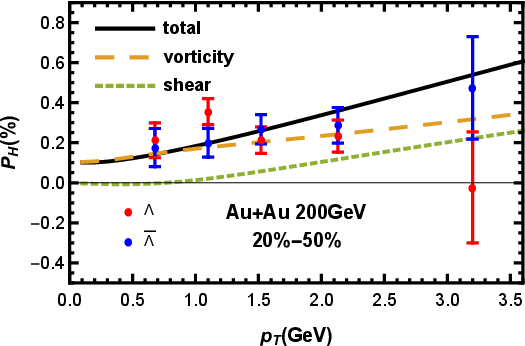}
\par\end{centering}
\caption{The results for $\left\langle P^{z}\sin(2\phi_{p})\right\rangle $
and $P_{H}$ as functions of $p_{T}$ in Au+Au collisions at 200 GeV.
We use the parameters of 30-40\% central collisions as an approximation.\label{fig:P_pT}}
\end{figure}

\begin{figure}
\begin{centering}
\includegraphics[scale=0.8]{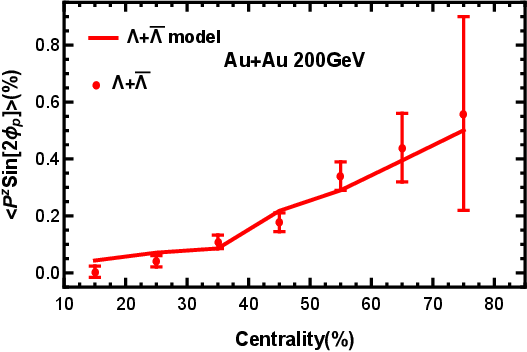} \includegraphics[scale=0.8]{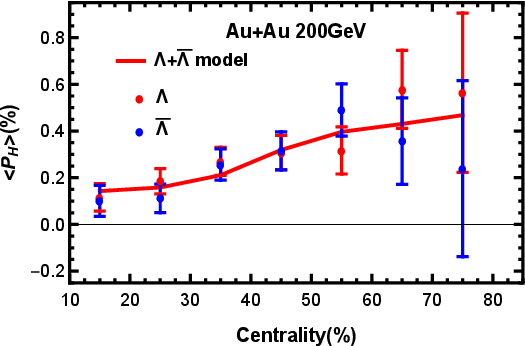}
\par\end{centering}
\caption{The result for the centrality dependence of $P_{z}$ and $P_{H}$
in Au+Au collisions at 200 GeV. \label{fig:Pz_centrality}}
\end{figure}


Also we calculated the transverse momentum dependence of polarziations
following the analytical formula in Eq. (\ref{eq:py-pz-pt}). The
results are shown in Fig. \ref{fig:P_pT}. For $\left\langle P^{z}\sin(2\phi_{p})\right\rangle $,
the experimental data are almost constant when $p_{T}\gtrsim1\text{ GeV}$
while the calculated result grows with increasing $p_{T}$. The vorticity
contribution is negative while the shear contribution is positive
and has the same sign as the experimental data, consistent with the
results by hydrodynamic models. For $P_{H}$, the vorticity contribution
is doniment at low $p_{T}$ and increases with growing $p_{T}$. The
shear contribution is approximately zero at $p_{T}\lesssim1$ GeV
and increases more rapidly but still smaller than the vorticity contribution.


\begin{table}
\begin{centering}
\begin{tabular}{|c|c|c|c|c|c|c|c|}
\hline 
centrality & $R$ (fm) & $T$ (MeV) & $\rho_{0}$ & $\rho_{2}$ & $\epsilon$ & $\alpha_{1}$ & $\tau_{f}$(fm/c)\tabularnewline
\hline 
\hline 
0-10\% & 13.8 & 91 & 1.35 & 0.016 & 0.035 & $-0.002$ & 12.0\tabularnewline
\hline 
10-20\% & 12.6 & 94 & 1.32 & 0.023 & 0.051 & $-0.002$ & 10.1\tabularnewline
\hline 
20-30\% & 11.5 & 97 & 1.28 & 0.032 & 0.0695 & $-0.002$ & 9.0\tabularnewline
\hline 
30-40\% & 10.0 & 101 & 1.23 & 0.043 & 0.08 & $-0.002$ & 6.6\tabularnewline
\hline 
40-50\% & 8.6 & 108 & 1.15 & 0.048 & 0.0935 & $-0.002$ & 4.3\tabularnewline
\hline 
50-60\% & 7.7 & 115 & 1.04 & 0.052 & 0.0985 & $-0.002$ & 3.38\tabularnewline
\hline 
60-70\% & 7.0 & 129 & 0.92 & 0.056 & 0.105 & $-0.002$ & 3.0\tabularnewline
\hline 
\end{tabular}
\par\end{centering}
\caption{Parameters in the solvable model that are used in this paper for Pb+Pb
collisions at $\sqrt{s_{NN}}=5.02$ TeV. \label{tab:para-Pb-Pb}}
\end{table}

\begin{figure}
\begin{centering}
\textcolor{red}{\includegraphics[scale=0.8]{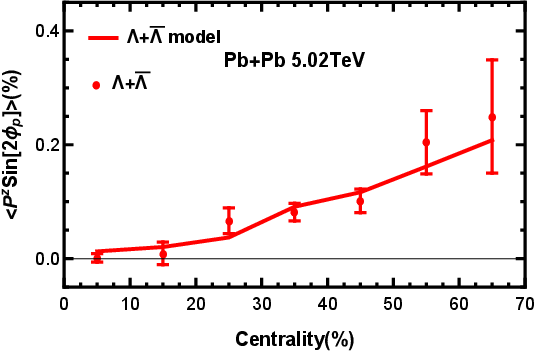}
\includegraphics[scale=0.8]{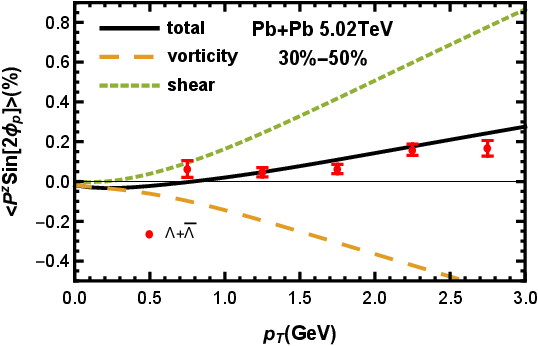}}
\par\end{centering}
\caption{Spin polarizations along the beam direction in Pb+Pb collisions at
5.02 TeV. The parameters for 30-40\% central collisions in Table \ref{tab:para-Pb-Pb}
are used in calculating the transverse momentum dependence of $\left\langle P^{z}\sin(2\phi_{p})\right\rangle $.
\label{fig:Pz-Pb}}
\end{figure}

\begin{figure}
\begin{centering}
\textcolor{red}{\includegraphics[scale=0.8]{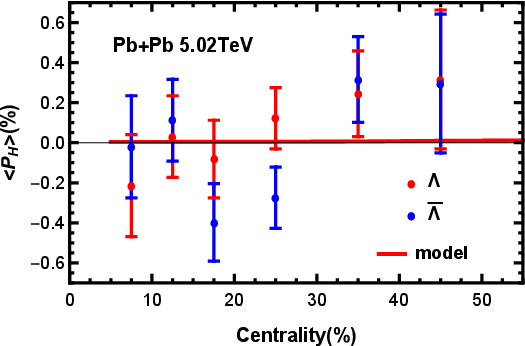}
\includegraphics[scale=0.8]{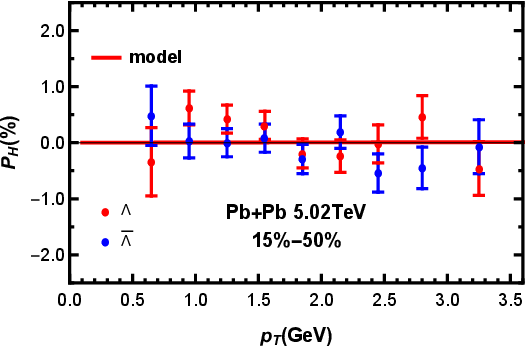}}
\par\end{centering}
\caption{Spin polarizations along the angular momentum direction in Pb+Pb collisions
at 5.02 TeV. The parameters for 20-30\% central collisions in Table
\ref{tab:para-Pb-Pb} are used in calculating the transverse momentum
dependence of $P_{H}$. \label{fig:Py-Pb}}
\end{figure}


Using the analytical formula in Eq. (\ref{eq:py-pz-total}), we can
calculate the centrality dependence of $P^{z}$ and $P_{H}$ in the
form of $\left\langle P^{z}\sin(2\phi_{p})\right\rangle $ and $\left\langle P_{H}\right\rangle $.
The comparasion of the theoretical result with data is shown in Fig.
\ref{fig:Pz_centrality}. One can see that the theoretical curve grows
from central to peripheral collisions which can describe the experimental
data. 


\subsection{Pb+Pb collisions at 5.02 TeV}

The parameters that we choose in the solvable model for Pb+Pb collisions
at $\sqrt{s_{NN}}=5.02$ TeV and different centralities are listed
in Table \ref{tab:para-Pb-Pb}. The parameter $\alpha_{1}$ is assumed
to be independent of the centrality \citep{STAR:2011hyh}. The freeze-out
time is approximated as 1.3 times $\tau_{f}$ at 200 GeV. Other parameters
are determined by fitting transverse momentum spectra \citep{ALICE:2019hno},
directed \citep{ALICE:2019sgg} and elliptic flows \citep{ALICE:2018yph}.
The calculated results for the spin polarization variable $\left\langle P^{z}\sin(2\phi_{p})\right\rangle $
in the beam direction as functions of centrality and transverse momentum
are shown in Fig. \ref{fig:Pz-Pb}, which successfully reproduces
experimental data \citep{ALICE:2021pzu}. The results for the spin
polarization in the angular momentum direction are almost zero as
shown in Fig. \ref{fig:Py-Pb}, which is consistent with experimental
data \citep{ALICE:2019onw}. 


\section{Summary}

We propose a solvable model for spin polarizations based on the blast-wave
picture of heavy-ion collisions with flow-momentum correspondence
at the leading order. To our knowledge, this is the first analytically
solvable model for spin polarizations in heavy-ion collisions. The
analytical solution we find has following features: (1) It not only
gives the exact azimutal angle dependences of spin polarizations in
the beam and angular momentum directions, but also gives their exact
transverse momentum dependences; (2) It has a symmetry between the
contribution from the vorticity and from the shear stress tensor;
(3) It can be improved order by order through expansion in $\delta\phi=\phi_{b}-\phi_{p}$
and $\delta\eta=\eta-Y$. The solvable model can describe almost all
available data for spin polarizations in the beam and angular momentum
directions with a few parameters constrained by transverse momentum
spectra and collective flows of hadrons.

\begin{acknowledgments}
The work was completed during the time when Q.W. visited the nuclear
theory group at McGill University as a visiting professor. Q.W. thanks
C. Gale and S. Jeon for their hospitality and thanks S. Jeon and X.-Y.
Wu for insightful discussions. The work is supported in part by the
National Natural Science Foundation of China (NSFC) under Grant Nos.
12135011, 12147101 and 12325507, the Strategic Priority Research Program
of the Chinese Academy of Sciences (CAS) under Grant No. XDB34030102,
the National Key Research and Development Program of China under Grant
No. 2022YFA1604900, and the Guangdong Major Project of Basic and Applied
Basic Research under Grant No. 2020B0301030008 (G.-L.M.).
\end{acknowledgments}

\bibliographystyle{unsrt}
\phantomsection\addcontentsline{toc}{section}{\refname}\bibliography{referances}

\end{document}